\documentclass[]{interact}

\usepackage{rotating}
\usepackage{array}
\usepackage{graphicx}
\usepackage{subcaption}

\usepackage[natbibapa]{apacite}
\bibpunct[, ]{(}{)}{;}{a}{,}{,}

\usepackage{amsfonts}

\usepackage{hyperref}
\hypersetup{
    colorlinks = true,
    linkcolor=black,
    anchorcolor=black,
    citecolor=black,
    filecolor=black,
    menucolor=black,
    runcolor=black,
    urlcolor=black
}

\makeatletter
\def\maxwidth{\ifdim\Gin@nat@width>\linewidth\linewidth
\else\Gin@nat@width\fi}
\makeatother
\let\Oldincludegraphics\includegraphics
\renewcommand{\includegraphics}[1]{\Oldincludegraphics[width=\maxwidth]{#1}}

\theoremstyle{plain}

\theoremstyle{definition}

\theoremstyle{remark}

\usepackage[T1]{fontenc}
\usepackage[utf8]{inputenc}

\usepackage{etoolbox}
\usepackage{tcolorbox}
\usepackage{array}

\tcbset{colback=gray!5!white,colframe=gray!85!black}

\AtBeginEnvironment{quote}{\vspace{3mm}\begin{tcolorbox}[leftrule=2mm,bottomrule=0mm,toprule=0mm,rightrule=0mm,boxsep=0mm,grow to right by=-3mm, grow to left by=-3mm]\small}
\AtEndEnvironment{quote}{ \end{tcolorbox}}

\author{
	\name{Wouter Groeneveld\textsuperscript{a}\thanks{CONTACT Wouter Groeneveld. Email: wouter@brainbaking.com}}
	\affil{\textsuperscript{a}Independent Researcher, Hasselt, Belgium}
}

\usepackage{color}
\usepackage{fancyvrb}

\DefineVerbatimEnvironment{Highlighting}{Verbatim}{commandchars=\\\{\}}
\usepackage{framed}
\definecolor{shadecolor}{RGB}{248,248,248}
\newenvironment{Shaded}{\begin{snugshade}}{\end{snugshade}}
\newcommand{\KeywordTok}[1]{\textcolor[rgb]{0.13,0.29,0.53}{\textbf{#1}}}

\newcommand{\BuiltInTok}[1]{#1}

\newcommand{\NormalTok}[1]{#1}
\begin{document}

\articletype{}

\title{Leveraging Creativity as a Problem Solving Tool in Software
Engineering}

\maketitle

\begin{abstract}
Today's software engineering (SE) complexities require a more diverse
tool set going beyond technical expertise to be able to successfully
tackle all challenges. Previous studies have indicated that creativity
is a prime indicator for overcoming these hurdles. In this paper, we
port results from creativity research in the field of cognitive
psychology to the field of SE. After all, programming is a highly
creative endeavour. We explore how to leverage creativity as a practical
problem solving tool to wield for software developers. The seven
distinct but intertwined creative problem solving themes unfolded in
this paper are accompanied with practical perspectives, specifically
geared for software professionals. Just like technical skills such as
knowledge of programming languages, we believe that creativity can be
learned and improved with practice.
\end{abstract}


This is the pre-print of a paper that has been accepted for publication
in IEEE Software - Special Issue on Creativity and Software Development.

\hypertarget{introduction}{%
\section{Introduction}\label{introduction}}

\label{sec:intro}

What is the difference between a good and a great software engineer when
it comes to problem solving? Solving intricate programming problems on
high level architectural and low level design levels requires mastery of
multiple domain general and domain specific hard and soft skills where
creativity comes up on top as one of the the most prevalent factors to
successfully facilitate tackling difficult problems
\citep{sedelmaier2014practicing}.

Creativity can be used as a form of self-expression, for example in
creative coding sessions aimed at exploring and learning. Instead, in
this paper, we focus on creativity as a practical tool to solve a
problem. By emphasizing on the right combination between creativity and
critical thinking, taking into account the context and constraints of
the problem, creativity can be wielded as a powerful tool---not just as
a goal in itself. In other words: \emph{creativity is the means, not the
goal}.

Yet what exactly is creativity? According to cognitive psychologists, an
idea is creative if it meets these descriptions: it is considered novel
and original; it is of high quality; and it is relevant to the task at
hand \citep{kaufman2007creativity}. However, this essentialists' view on
creativity does have its shortcomings, for example, by completely
ignoring context. Recent advances in creativity research showcase a
gradual shift towards a systemic approach where creativity can be seen
as a social verdict \citep{glaveanu2020advancing}, meaning your peers
decide whether or not your coding solution is considered creative.

\begin{figure}[h!]
    \centering
    \includegraphics{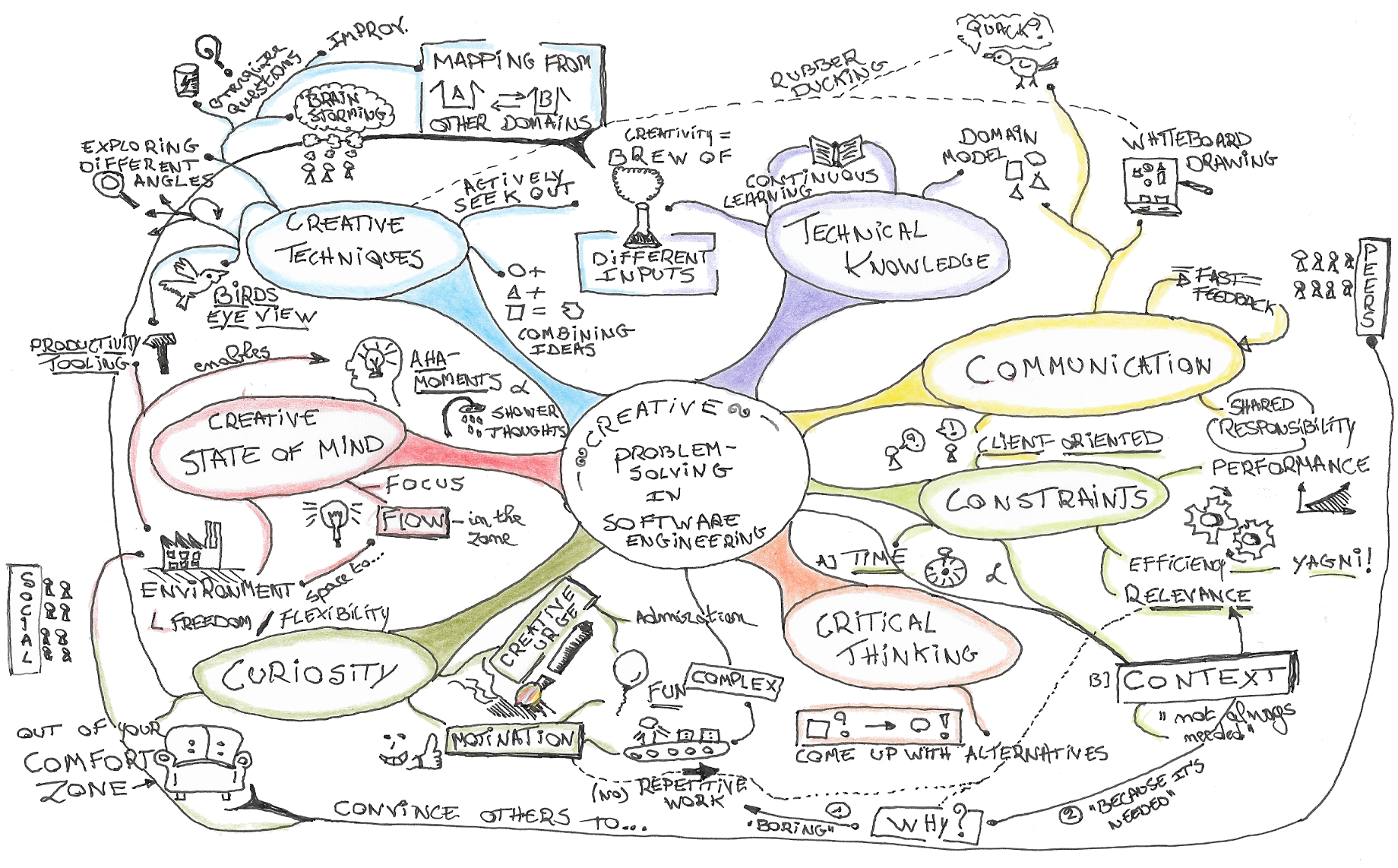}
    \caption{A mind map that summarizes identified themes on creative problem solving in SE (cited from paper). \label{fig:mindmap}}
 \end{figure}

This may sound like we have zero control of our own creative outcome and
are at the mercy of our colleagues when it comes to achieving the label
``creative''. Luckily, interviews and Delphi studies conducted by
\citet{groeneveld2021exploring} identified several distinct but highly
intertwined domains of creative problem solving in context of SE: (1)
\emph{technical knowledge}, (2) \emph{collaboration}, (3)
\emph{constraints}, (4) \emph{critical thinking}, (5) \emph{curiosity},
(6) \emph{a creative state of mind}, and lastly (7) practical
\emph{creative techniques}. A summary is available in Figure
\ref{fig:mindmap}.

This paper is part of a bigger project on exploring creativity in SE by
\citet{groeneveld2021exploring} who initially published above mind map.
Here, we provide a new practical view that translates the theory from
previous publications into practical tools for software engineers. We
believe that by taking advantage of these ideas, practitioners will be
able to get unstuck when facing a complex problem. In essence,
creativity can be leveraged as a problem solving tool in SE.

The remainder of this paper unfolds these seven creative problem solving
domains, providing new practical pointers on how to exploit them to help
us become more creative programmers.

\hypertarget{technical-knowledge}{%
\section{Technical Knowledge}\label{technical-knowledge}}

\label{sec:tech}

Can you approach a problem creatively if you do not have the necessary
technical domain knowledge to do so? For example, consider this Go
example code snippet:

\begin{Shaded}
\begin{Highlighting}[]
\KeywordTok{func}\NormalTok{ ToChannel(done \textless{}{-}}\KeywordTok{chan} \KeywordTok{struct}\NormalTok{\{\}) \textless{}{-}}\KeywordTok{chan}\NormalTok{ any \{}
\NormalTok{    c := }\BuiltInTok{make}\NormalTok{(}\KeywordTok{chan}\NormalTok{ any)}
    \KeywordTok{go} \KeywordTok{func}\NormalTok{() \{}
       \KeywordTok{defer} \BuiltInTok{close}\NormalTok{(c)}
       \KeywordTok{select}\NormalTok{ \{}
       \KeywordTok{case}\NormalTok{ \textless{}{-}done:}
          \KeywordTok{return}
\NormalTok{       \}}
\NormalTok{    \}()}
    \KeywordTok{return}\NormalTok{ c}
\NormalTok{\}}
\end{Highlighting}
\end{Shaded}

Without some base level of Go's asynchronous channel concepts, it would
be very challenging to refactor the above piece of code, let alone
borrow channel concept to creatively deploy them in another language
without that feature baked into it. And yet, that is exactly what many
expert practitioners do and what their peers call creative, according to
our interviews \citep{groeneveld2021exploring}.

Consider the birth of the Kotlin programming language. The engineers who
designed the language inspected and heavily borrowed concepts from
existing programming languages. Kotlin features concepts of Java
(classes, autoboxing, runtime safety guarantees, etc.), Scala (primary
constructors, the \texttt{val} keyword, etc.), C\# (ideas of
\texttt{get/set} properties and extensions, etc.), and Groovy (the
\texttt{it} shorthand)\footnote{For more information, see Andrey
  Breslav's \emph{Shoulders of Giants: Languages Kotlin Learned From}
  talk at GeekOUT 2018 at https://youtu.be/Ljr66Bg--1M.}.

In one of our interviews while exploring the role of creativity in SE, a
participant summarized his thoughts \citep{groeneveld2021exploring}:

\begin{quote}
The bottom line is that creativity is the brew of different inputs---and
usually, I actively look up those inputs.
\end{quote}

No input, no (creative) output. Yet gathering input alone is not
sufficient: research in the field of cognitive psychology has shown that
for productivity levels to increase, one also has to internalize
knowledge and act upon it \citep{csikszentmihalyi1997flow}.

In sum, we have to cultivate a Personal Knowledge Management tool where
findings summarized in our own words reside that can be connected to
form new and exciting creative ideas. The aim here is to evolve loosely
coupled information to tightly coupled knowledge and (new) insight. This
is not only applicable for academics publishing papers, but also for
developers publishing code, mastering languages and, and writing
technical articles.

\hypertarget{collaboration}{%
\section{Collaboration}\label{collaboration}}

\label{sec:coll}

Does creativity exist in isolation? Collaborative teamwork---ideally, a
heterogeneous group of people with different backgrounds and expertise
sets---might end up radically changing the field they work in. For
example, consider the mid-2000s Extreme Tuesday Club hosted by expert
developers in London. The Club turned out to be a great breeding ground
for several important and established SE techniques such as Continuous
Delivery, Kanban, unit test mocking frameworks, and more. Additionally,
other like-minded people in SE successfully mimicked the concept of the
informal gathering, resulting in software craftsmanship meetups in every
conceivable big city.

\begin{figure}[h!]
  \centering
  \includegraphics{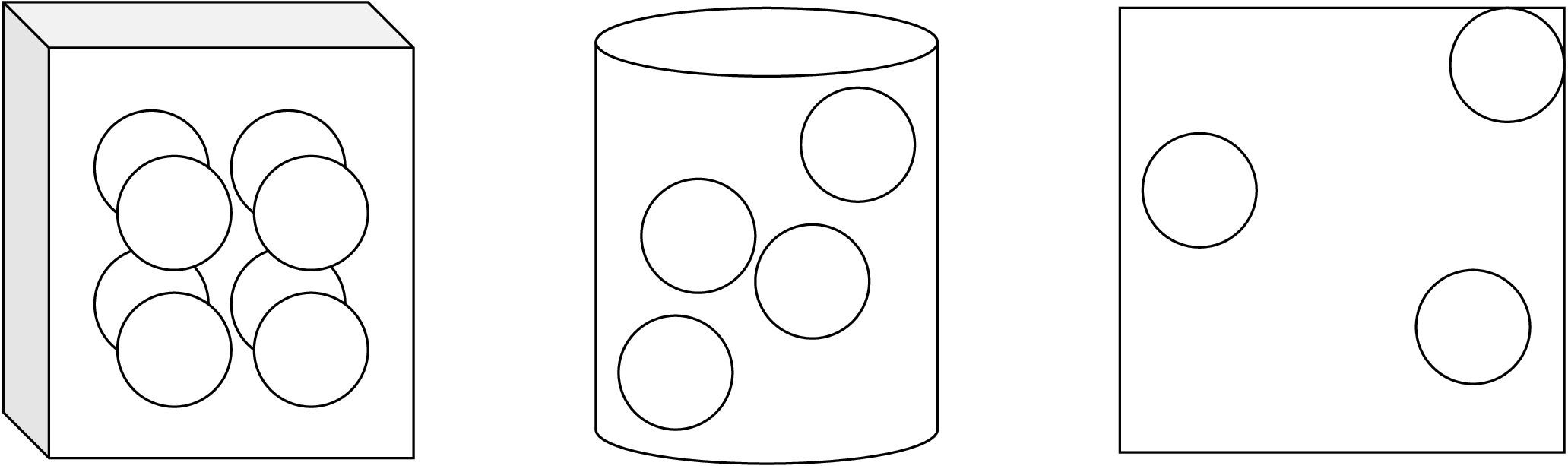}
  \caption{A schematic of the liquid network principle. Left: solid matter. Ideas set in stone due to the definite shape and volume. Middle: liquid matter. Allowing the creation of new connections. Right: gas matter. Too volatile to make ideas stick. \label{fig:liquid}}
\end{figure}

Contemporary SE heavily relies on teamwork. According to
\citet{csikszentmihalyi1997flow}, the creative power of the individual
is negligible. Instead, organisations should focus on engineering
pockets of creative collaboration, or what \citet{johnson2011good} calls
\emph{liquid networks}. Solid matter is tightly clustered together and
allowed little wiggle room, while gas matter is too volatile to make
ideas really stick. Instead, Johnson suggests we focus on creating a
liquid network, as visible in the middle of Figure \ref{fig:liquid}: one
where ideas can bounce around and form new ones without risking to
become too unstable. While Johnson's work is yet to be studied in SE,
networks like the Extreme Tuesday Club proves the concept shows promise.

We are more creative when in the (virtual) vicinity of creative
coworkers, yet at the same time, sometimes those very coworkers heavily
impede our creative flow. Social issues in and around software
development teams severely affects team performance.
\citet{tamburri2016architect} call this phenomena \emph{social debt},
and just like the well-known \emph{technical debt} describing the
implied cost of future code changes that comes with \emph{code smells},
social debt has its own anti-patterns called \emph{community smells}.

The following selection of community smells as identified by Tamburri et
al.~demonstrates the principle and its destructive relation with
creativity more clearly:

\begin{itemize}

\item
  \emph{Lone Wolf}---a person who does not take others' opinions into
  consideration and ignores rules set out by the team;
\item
  \emph{Black Cloud}---information overload and no way to manage
  information, resulting in the potential loss of creative ideas;
\item
  \emph{Hyper-Community}---a too volatile environment where no idea
  truly sticks or no time is given to thoroughly evaluate ideas;
\item
  \emph{Cookbook Development}---developers stuck in their ways refusing
  to adapt to new technologies who are hostile towards new ideas.
\end{itemize}

A few dangerous combinations of the above smells will have a devastating
impact on team morale, and thus, creativity. While openly and critically
discussing ideas is good, an excess of conflicts is detrimental
creativity. The transition from code smell to community smell does not
require a big leap of faith: many code smells find their root in
community smells.

\hypertarget{constraints}{%
\section{Constraints}\label{constraints}}

\label{sec:constr}

\begin{figure}[h!]
     \centering
     \begin{subfigure}[b]{0.48\textwidth}
         \centering
         \includegraphics{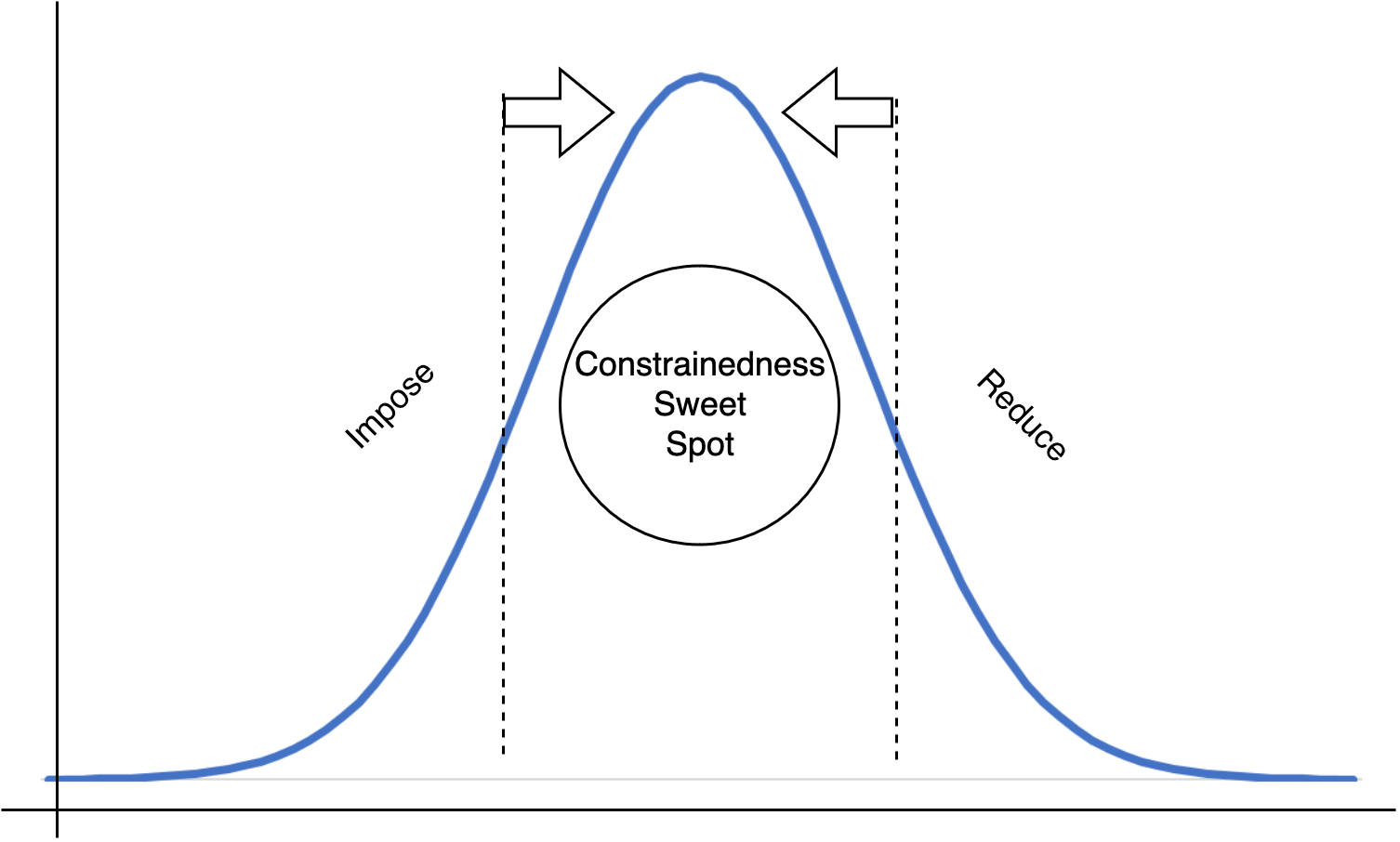}
         \caption{The constrainedness sweet spot.}
     \end{subfigure}
     \hfill
     \begin{subfigure}[b]{0.48\textwidth}
         \centering
         \includegraphics{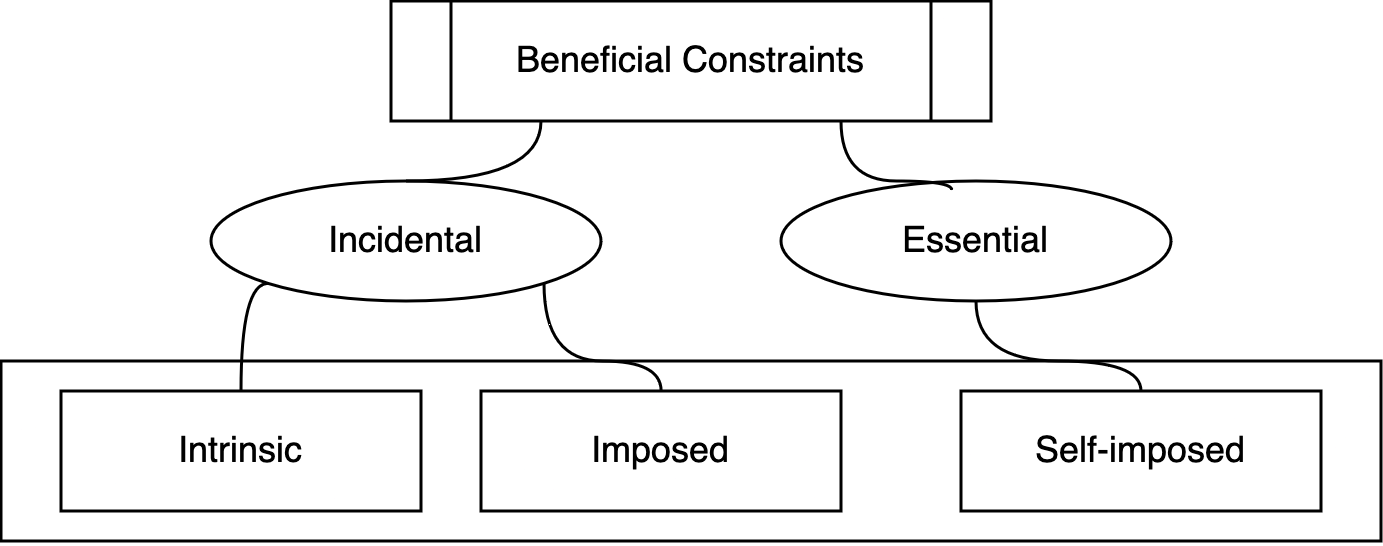}
         \caption{The taxonomy of beneficial constraints.}
     \end{subfigure}
        \caption{Behaviour and taxonomies of constraints (own diagrams).}
        \label{fig:constraints}
\end{figure}

Can we readjust the constraints of a software development project in
such a way that it increases the creative thinking throughput?
Constraints do drastically improve creativity---up to a point. Too
little and the endless sea of freedom indefinitely postpones our
motivation to act. Too much and the stress level diminishes our creative
thinking capabilities. \citet{biskjaer2020task} suggest we aim for a
sweet spot in-between both, as visible in Figure \ref{fig:constraints}a,
to either impose new ones when we are feeling bored or try to reduce
constraints when we feel stressed.

But before we can do that, we first have to identify different
categories of constraints and how they manifest in SE.
\citet{elster2000ulysses} defines a taxonomy of constraints, as visible
in Figure \ref{fig:constraints}b. The sweet spot and taxonomy concepts
are domain-general, requiring further empirical evidence in SE, but our
previous work suggests they do translate well into SE.

First, \emph{intrinsic constraints} are inherent to the problem at hand.
If you are a programmer, you will have to work with the software and
hardware that is currently available to you. If you wrote programs in
the early nineties, chances of bumping against the hardware limit are
much higher than nowadays machines, yet these tight constraints were
also a source of inspiration. For instance, the Game Boy (GB) game
\emph{X} boasts a first-person true 3D polygonal rendered flight scene,
while in 1992 was technically impossible to do 3D on the GB that on
paper only supports 2D sprites. The developers managed to successfully
bend those intrinsic hardware constraints to their advantage.

Second, \emph{imposed constraints} are constraints handed out by a
stakeholder. For example, typical budget or time constraints as agreed
upon by customers and management are part of this category. While the
difference between intrinsic and imposed constraints is subtle---they
are both grouped together as \emph{incidental} constraints---there is a
big psychological difference. We would rather accept contemporary
hardware constraints than constraints we were told to adhere to.
Successfully working with and transforming legacy code falls within this
category. Modern software development techniques applied to older
programming environments are the prime example, such as
\emph{Test4z}\footnote{See https://mainframe.broadcom.com/test4z/.} that
enables unit testing for mainframes.

Third, \emph{self-imposed constraints} are constraints we add ourselves
to deliberately push our creative problem solving skills to the limit.
Limitations can be seen as a guide to your creative solutions. Many game
developers are still creating new games for older hardware because these
tight constraints force them to come out of their comfort zone.
Purposely writing methods with three or less arguments and twenty or
less lines of code to force yourself to think about responsibilities and
clean code is another example of self-imposed constraints, as is the
philosophy and deliberately simple design of the Go programming
language.

\hypertarget{critical-thinking}{%
\section{Critical Thinking}\label{critical-thinking}}

\label{sec:crit}

When are software developers not so creative? According to the
practitioners we interviewed, when they blindly take over answers from
AI-powered code generation tools, when alternatives are not being
considered, when just their tasks are being ticked off, and when
repetition kicks in. A participant elaborates:

\begin{quote}
{[}I am not creative when{]} I just try it until it works, such as
fixing dependencies.
\end{quote}

Many interviewees emphasize the importance of critically evaluating and
iterating on ideas. This \emph{verification} step perfectly matches one
of the five stages of the creative process as first identified by Graham
Wallas \citep{sadler2015wallas}:

\begin{enumerate}
\def\labelenumi{\arabic{enumi}.}

\item
  \emph{Participate}---the bulk of the work happens here;
\item
  \emph{Incubate}---the creator takes a bit of distance from their work
  and interrupts the process to facilitate insight;
\item
  \emph{Illuminate}---The ``aha''-moment (see Section
  \ref{sec:stateofmind}) that will not happen without the sweat of the
  \emph{participate} step;
\item
  \emph{Verify}---critically evaluating and iterating on ideas;
\item
  \emph{Present/accept}---the solution is only creative if your peers
  accept it (see Section \ref{sec:intro}).
\end{enumerate}

This process is recursive where any step (except for final step 5) might
jump to any other step. Csikszentmihalyi suggests that taking a step
back to critically evaluate your creation is one of the key aspects to
creative success. For a software developer, this is equally important as
they rely on (automated) tooling to facilitate the iterative evaluation
step: unit tests that turn red or green in the pipeline after each
commit, fitness tests that pass or fail to signal architectural
problems, etc.

Skimping on the verification step leads to critical thinking falacies
that in SE often result in common coding mistakes.
\citet{hermans2021programmer} collected many of these mistakes based on
interviews and empirical studies in the SE industry. For example, many
developers rely on \emph{transfer during learning} to pick up a new
language: type inference in Kotlin
(\texttt{val\ myArr\ =\ arrayOf(1,\ 2)}) helps reducing the jump to
JavaScript's dynamic duck typing (\texttt{let\ myArr\ =\ {[}1,\ 2{]}}).
However, that transfer can also lead to \emph{cross-language
interference}. The usage of the \texttt{const} keyword in Kotlin creates
compile-time constants to primitives and strings. In JavaScript,
\texttt{const} defines a runtime constant reference to a value: you can
still push new values to a \texttt{const} array.

\hypertarget{curiosity}{%
\section{Curiosity}\label{curiosity}}

\label{sec:curio}

Can we motivate software developers to become more curious, and thus in
turn more creative? Research by Csikszentmihalyi reveals that curiosity
and perseverance play a major role in creative success, while
\citet{gross2020cultivating} highlight that although empirical evidence
for the curiosity-creativity connection is scarce, curiosity may indeed
benefit creativity.

Experiments in SE education also suggest that creative students get out
of their comfort zone more often than their less-creative peers
\citep{groeneveld2022undergraduate}. In our own interviews exploring the
role of creativity in SE, curiosity is often cited as a core proponent
of creativity. A participant found his colleague to be more creative
than himself because they were frequently reading new books, reading
articles, and trying out new programming languages.

Without staying curious, less new technical knowledge will be gained
(Section \ref{sec:tech}), decreasing the chance of idea forming and
merging. Without perseverance, imposed constraints will start to weigh
(too) heavy (Section \ref{sec:constr}) and the iterative creative
process from Section \ref{sec:crit} will yield less polished results.

Many software developers we interviewed keep themselves motivated by
what they call ``fooling around'' in their free time: discovering new
languages, frameworks, and techniques by programming playful small
diversions. They are having fun and learn new things at the same time,
not necessarily leveraging creativity to solve a problem at hand, but
using their curiosity to further feed their creative craving. Employers
can further support this by organizing reading groups, hackatons, and
knowledge sharing afternoons. In the ever changing landscape of SE,
continuous learning has become critical to effectively work,
communicate, and innovate on an international scale.

\hypertarget{creative-state-of-mind}{%
\section{Creative State of Mind}\label{creative-state-of-mind}}

\label{sec:stateofmind}

How can software developers generate more insights? As mentioned in
Section \ref{sec:crit}, no Eureka effect without ample blood, sweat, and
tears. Yet getting in the right creative mood does wonders for the
frequency of new ideas that ``suddenly'' pop up in our minds. Again, a
great body of research work has been published by Mihaly
Csikszentmihalyi on how to achieve a state of what he calls flow or the
\emph{optimal experience} \citep{csikszentmihalyi1997flow}.

In order for this flow state to occur, just the right amount of
challenge combined with skill level is required. Too challenging, and we
experience anxiety. Too little challenge and too high skill level, we
will feel bored. When we interviewed software engineers on the
requirements to be creative, they answered in vaguer terms such as
\emph{if everything feels right} and \emph{the feeling that you're
really in the {[}flow{]} zone}, hinting at Csikszentmihalyi's optimal
experience.

According to our SE respondents, the environment plays a major role in
facilitating flow. Some jump in their car to think while others take a
shower or go for a run. By not actively thinking about the problem, they
\emph{are} actively thinking about the problem: this is the incubate
step of Section \ref{sec:crit}.

Corporate environments or home offices of software professionals should
provide the necessary flexibility and freedom to help keep the flow
flowing, even though workspace and decorations are highly subjective.
\citet{thoring2019creative} compiled a list of unique space types that
should be present in each work environment to maximize creative
potential:

\begin{itemize}

\item
  A private space to focus;
\item
  A collaborative and/or making/experimentation space;
\item
  An exhibition and/or presentation/sharing space;
\item
  A relaxation space;
\item
  An unusual space that might trigger insight;
\item
  An incubation and reflection space.
\end{itemize}

Strictly separating these spaces and their temporary inhabitants is
important: excessive interrupts from your private focus space can
negatively impact the creative process, especially for information
workers.

\hypertarget{creative-techniques}{%
\section{Creative Techniques}\label{creative-techniques}}

Creative SE work is boosted by employing various practical tools and
techniques, such as the aforementioned analogy technique (mapping
solutions from another domain; Section \ref{sec:tech}), or actively
seeking out feedback, both self-reflective and external (Section
\ref{sec:coll}). Our interviewees regularly mentioned these techniques
in context of problem solving.

A more thorough exploration of the usage of creative techniques in SE
was conducted by \citet{bobkowska2019exploration} using a
training-application-feedback cycle. According to the authors, the best
creative results were achieved by leveraging a mix of these techniques.
The identified techniques are grouped into four themes: (1)
\emph{interpersonal skills} (see Section \ref{sec:coll}), (2)
\emph{creativity skills} such as associative thinking, (3)
\emph{motivational skills}, and (4) \emph{overcoming obstacles}. The
following selection illustrates some of the techniques:

\begin{itemize}

\item
  \emph{Let's Invite Them}: what if we invite for example Joshua Bloch,
  one of the lead architects of the Java platform---what would he
  suggest we do?;
\item
  \emph{What If\ldots{}}: what if we did not need distributed
  transactions---how would we tackle the problem then?;
\item
  \emph{Reverse Brainstorming}: what don't we like about our current
  distributed transaction implementation?;
\item
  \emph{I Could Be More Creative If\ldots{}}: what if the organisation
  encouraged a more flexible work plan to facilitate flow?
\end{itemize}

These thought experiments can be conducted individually or in group.

\hypertarget{conclusion}{%
\section{Conclusion}\label{conclusion}}

\label{sec:conclusion}

In this paper, we zoomed in on a systematic view of creative problem
solving in SE that consists out of seven distinct but heavily
intertwined domains: \emph{technical knowledge}, \emph{collaboration},
\emph{constraints}, \emph{critical thinking}, \emph{curiosity},
cultivating \emph{a creative state of mind}, and practical
\emph{creative techniques}.

Software practitioners can overcome difficult problems that typically
arise when developing complex software systems by wielding and combining
the tools as discussed. For example, approach sitations from different
angles by adding more constraints or taking a few away, and when stuck.
Be on the lookout for new input to get that creative brew going.
Stimulate or introduce other knowledge domains that might generate
original approaches to tackle the problem. Be mindful of interruptions
and agree on how these should be tackled. Integrate moments of playful
discovery in your professional life. Identify and explicitly discuss
social debt next to technical debt in retrospectives.

But most of all, remind your SE colleagues that creativity is an
attainable skill: just like programming, it can be learned with practice
and patience.

\hypertarget{author-biography}{%
\section{Author Biography}\label{author-biography}}

Dr.~Wouter Groeneveld is an independent software architect and
researcher based in Hasselt, Belgium. In 2018, after spending more than
a decade as a software engineer in the industry, Wouter returned to
academia by joining the Department of Computer Science at KU Leuven to
combine teaching with researching creativity in SE education. In 2023,
he published \emph{The Creative Programmer}, a practical book
summarizing his academic research to help bridge the gap between
academia and industry. After obtaining his PhD, Wouter made the switch
back to the industry.

\bibliography{report}

\begin{thebibliography}{}

\bibitem [\protect \citeauthoryear {%
Biskjaer%
\ \protect \BOthers {.}}{%
Biskjaer%
\ \protect \BOthers {.}}{%
{\protect \APACyear {2020}}%
}]{%
biskjaer2020task}
\APACinsertmetastar {%
biskjaer2020task}%
\begin{APACrefauthors}%
Biskjaer, M\BPBI M.%
, Christensen, B\BPBI T.%
, Friis-Olivarius, M.%
, Abildgaard, S\BPBI J.%
, Lundqvist, C.%
\BCBL {}\ \BBA {} Halskov, K.%
\end{APACrefauthors}%
\unskip\
\newblock
\APACrefYearMonthDay{2020}{}{}.
\newblock
{\BBOQ}\APACrefatitle {How task constraints affect inspiration search
  strategies} {How task constraints affect inspiration search
  strategies}.{\BBCQ}
\newblock
\APACjournalVolNumPages{International Journal of Technology and Design
  Education}{30}{1}{101--125}.
\PrintBackRefs{\CurrentBib}

\bibitem [\protect \citeauthoryear {%
Bobkowska%
}{%
Bobkowska%
}{%
{\protect \APACyear {2019}}%
}]{%
bobkowska2019exploration}
\APACinsertmetastar {%
bobkowska2019exploration}%
\begin{APACrefauthors}%
Bobkowska, A\BPBI E.%
\end{APACrefauthors}%
\unskip\
\newblock
\APACrefYearMonthDay{2019}{}{}.
\newblock
{\BBOQ}\APACrefatitle {Exploration of creativity techniques in software
  engineering in training-application-feedback cycle} {Exploration of
  creativity techniques in software engineering in
  training-application-feedback cycle}.{\BBCQ}
\newblock
\BIn{} \APACrefbtitle {Enterprise and Organizational Modeling and Simulation:
  15th International Workshop, EOMAS 2019, Held at CAiSE 2019, Rome, Italy,
  June 3--4, 2019, Selected Papers 15} {Enterprise and organizational modeling
  and simulation: 15th international workshop, eomas 2019, held at caise 2019,
  rome, italy, june 3--4, 2019, selected papers 15}\ (\BPGS\ 99--118).
\PrintBackRefs{\CurrentBib}

\bibitem [\protect \citeauthoryear {%
Csikszentmihalyi%
}{%
Csikszentmihalyi%
}{%
{\protect \APACyear {1997}}%
}]{%
csikszentmihalyi1997flow}
\APACinsertmetastar {%
csikszentmihalyi1997flow}%
\begin{APACrefauthors}%
Csikszentmihalyi, M.%
\end{APACrefauthors}%
\unskip\
\newblock
\APACrefYearMonthDay{1997}{}{}.
\newblock
{\BBOQ}\APACrefatitle {Flow and the psychology of discovery and invention}
  {Flow and the psychology of discovery and invention}.{\BBCQ}
\newblock
\APACjournalVolNumPages{HarperPerennial, New York}{39}{}{}.
\PrintBackRefs{\CurrentBib}

\bibitem [\protect \citeauthoryear {%
Elster%
}{%
Elster%
}{%
{\protect \APACyear {2000}}%
}]{%
elster2000ulysses}
\APACinsertmetastar {%
elster2000ulysses}%
\begin{APACrefauthors}%
Elster, J.%
\end{APACrefauthors}%
\unskip\
\newblock
\APACrefYear{2000}.
\newblock
\APACrefbtitle {Ulysses unbound: Studies in rationality, precommitment, and
  constraints} {Ulysses unbound: Studies in rationality, precommitment, and
  constraints}.
\newblock
\APACaddressPublisher{}{Cambridge University Press}.
\PrintBackRefs{\CurrentBib}

\bibitem [\protect \citeauthoryear {%
Glaveanu%
\ \protect \BOthers {.}}{%
Glaveanu%
\ \protect \BOthers {.}}{%
{\protect \APACyear {2020}}%
}]{%
glaveanu2020advancing}
\APACinsertmetastar {%
glaveanu2020advancing}%
\begin{APACrefauthors}%
Glaveanu, V\BPBI P.%
, Hanchett~Hanson, M.%
, Baer, J.%
, Barbot, B.%
, Clapp, E\BPBI P.%
, Corazza, G\BPBI E.%
\BDBL {}others%
\end{APACrefauthors}%
\unskip\
\newblock
\APACrefYearMonthDay{2020}{}{}.
\newblock
{\BBOQ}\APACrefatitle {Advancing creativity theory and research: A
  socio-cultural manifesto} {Advancing creativity theory and research: A
  socio-cultural manifesto}.{\BBCQ}
\newblock
\APACjournalVolNumPages{The Journal of Creative Behavior}{54}{3}{741--745}.
\PrintBackRefs{\CurrentBib}

\bibitem [\protect \citeauthoryear {%
Groeneveld%
, Luyten%
, Vennekens%
\BCBL {}\ \BBA {} Aerts%
}{%
Groeneveld%
\ \protect \BOthers {.}}{%
{\protect \APACyear {2021}}%
}]{%
groeneveld2021exploring}
\APACinsertmetastar {%
groeneveld2021exploring}%
\begin{APACrefauthors}%
Groeneveld, W.%
, Luyten, L.%
, Vennekens, J.%
\BCBL {}\ \BBA {} Aerts, K.%
\end{APACrefauthors}%
\unskip\
\newblock
\APACrefYearMonthDay{2021}{}{}.
\newblock
{\BBOQ}\APACrefatitle {Exploring the Role of Creativity in Software
  Engineering} {Exploring the role of creativity in software
  engineering}.{\BBCQ}
\newblock
\BIn{} \APACrefbtitle {2021 IEEE/ACM 43nd International Conference on Software
  Engineering: Software Engineering in Society (ICSE-SEIS).} {2021 ieee/acm
  43nd international conference on software engineering: Software engineering
  in society (icse-seis).}
\PrintBackRefs{\CurrentBib}

\bibitem [\protect \citeauthoryear {%
Groeneveld%
, Martin%
, Poncelet%
\BCBL {}\ \BBA {} Aerts%
}{%
Groeneveld%
\ \protect \BOthers {.}}{%
{\protect \APACyear {2022}}%
}]{%
groeneveld2022undergraduate}
\APACinsertmetastar {%
groeneveld2022undergraduate}%
\begin{APACrefauthors}%
Groeneveld, W.%
, Martin, D.%
, Poncelet, T.%
\BCBL {}\ \BBA {} Aerts, K.%
\end{APACrefauthors}%
\unskip\
\newblock
\APACrefYearMonthDay{2022}{}{}.
\newblock
{\BBOQ}\APACrefatitle {Are Undergraduate Creative Coders Clean Coders? A
  Correlation Study} {Are undergraduate creative coders clean coders? a
  correlation study}.{\BBCQ}
\newblock
\BIn{} \APACrefbtitle {Proceedings of the 53rd ACM Technical Symposium on
  Computer Science Education V. 1} {Proceedings of the 53rd acm technical
  symposium on computer science education v. 1}\ (\BPGS\ 314--320).
\PrintBackRefs{\CurrentBib}

\bibitem [\protect \citeauthoryear {%
Gross%
, Zedelius%
\BCBL {}\ \BBA {} Schooler%
}{%
Gross%
\ \protect \BOthers {.}}{%
{\protect \APACyear {2020}}%
}]{%
gross2020cultivating}
\APACinsertmetastar {%
gross2020cultivating}%
\begin{APACrefauthors}%
Gross, M\BPBI E.%
, Zedelius, C\BPBI M.%
\BCBL {}\ \BBA {} Schooler, J\BPBI W.%
\end{APACrefauthors}%
\unskip\
\newblock
\APACrefYearMonthDay{2020}{}{}.
\newblock
{\BBOQ}\APACrefatitle {Cultivating an understanding of curiosity as a seed for
  creativity} {Cultivating an understanding of curiosity as a seed for
  creativity}.{\BBCQ}
\newblock
\APACjournalVolNumPages{Current Opinion in Behavioral Sciences}{35}{}{77--82}.
\PrintBackRefs{\CurrentBib}

\bibitem [\protect \citeauthoryear {%
Hermans%
}{%
Hermans%
}{%
{\protect \APACyear {2021}}%
}]{%
hermans2021programmer}
\APACinsertmetastar {%
hermans2021programmer}%
\begin{APACrefauthors}%
Hermans, F.%
\end{APACrefauthors}%
\unskip\
\newblock
\APACrefYear{2021}.
\newblock
\APACrefbtitle {The Programmer's Brain: What every programmer needs to know
  about cognition} {The programmer's brain: What every programmer needs to know
  about cognition}.
\newblock
\APACaddressPublisher{}{Manning}.
\PrintBackRefs{\CurrentBib}

\bibitem [\protect \citeauthoryear {%
Johnson%
}{%
Johnson%
}{%
{\protect \APACyear {2011}}%
}]{%
johnson2011good}
\APACinsertmetastar {%
johnson2011good}%
\begin{APACrefauthors}%
Johnson, S.%
\end{APACrefauthors}%
\unskip\
\newblock
\APACrefYear{2011}.
\newblock
\APACrefbtitle {Where good ideas come from: The natural history of innovation}
  {Where good ideas come from: The natural history of innovation}.
\newblock
\APACaddressPublisher{}{Penguin}.
\PrintBackRefs{\CurrentBib}

\bibitem [\protect \citeauthoryear {%
Kaufman%
\ \BBA {} Sternberg%
}{%
Kaufman%
\ \BBA {} Sternberg%
}{%
{\protect \APACyear {2007}}%
}]{%
kaufman2007creativity}
\APACinsertmetastar {%
kaufman2007creativity}%
\begin{APACrefauthors}%
Kaufman, J\BPBI C.%
\BCBT {}\ \BBA {} Sternberg, R\BPBI J.%
\end{APACrefauthors}%
\unskip\
\newblock
\APACrefYearMonthDay{2007}{}{}.
\newblock
{\BBOQ}\APACrefatitle {Creativity} {Creativity}.{\BBCQ}
\newblock
\APACjournalVolNumPages{Change: The Magazine of Higher
  Learning}{39}{4}{55--60}.
\PrintBackRefs{\CurrentBib}

\bibitem [\protect \citeauthoryear {%
Sadler-Smith%
}{%
Sadler-Smith%
}{%
{\protect \APACyear {2015}}%
}]{%
sadler2015wallas}
\APACinsertmetastar {%
sadler2015wallas}%
\begin{APACrefauthors}%
Sadler-Smith, E.%
\end{APACrefauthors}%
\unskip\
\newblock
\APACrefYearMonthDay{2015}{}{}.
\newblock
{\BBOQ}\APACrefatitle {Wallas’ four-stage model of the creative process: More
  than meets the eye?} {Wallas’ four-stage model of the creative process:
  More than meets the eye?}{\BBCQ}
\newblock
\APACjournalVolNumPages{Creativity research journal}{27}{4}{342--352}.
\PrintBackRefs{\CurrentBib}

\bibitem [\protect \citeauthoryear {%
Sedelmaier%
\ \BBA {} Landes%
}{%
Sedelmaier%
\ \BBA {} Landes%
}{%
{\protect \APACyear {2014}}%
}]{%
sedelmaier2014practicing}
\APACinsertmetastar {%
sedelmaier2014practicing}%
\begin{APACrefauthors}%
Sedelmaier, Y.%
\BCBT {}\ \BBA {} Landes, D.%
\end{APACrefauthors}%
\unskip\
\newblock
\APACrefYearMonthDay{2014}{}{}.
\newblock
{\BBOQ}\APACrefatitle {Practicing soft skills in software engineering: A
  project-based didactical approach} {Practicing soft skills in software
  engineering: A project-based didactical approach}.{\BBCQ}
\newblock
\BIn{} \APACrefbtitle {Overcoming Challenges in Software Engineering Education:
  Delivering Non-Technical Knowledge and Skills} {Overcoming challenges in
  software engineering education: Delivering non-technical knowledge and
  skills}\ (\BPGS\ 161--179).
\newblock
\APACaddressPublisher{}{IGI Global}.
\PrintBackRefs{\CurrentBib}

\bibitem [\protect \citeauthoryear {%
Tamburri%
, Kazman%
\BCBL {}\ \BBA {} Fahimi%
}{%
Tamburri%
\ \protect \BOthers {.}}{%
{\protect \APACyear {2016}}%
}]{%
tamburri2016architect}
\APACinsertmetastar {%
tamburri2016architect}%
\begin{APACrefauthors}%
Tamburri, D\BPBI A.%
, Kazman, R.%
\BCBL {}\ \BBA {} Fahimi, H.%
\end{APACrefauthors}%
\unskip\
\newblock
\APACrefYearMonthDay{2016}{}{}.
\newblock
{\BBOQ}\APACrefatitle {The architect's role in community shepherding} {The
  architect's role in community shepherding}.{\BBCQ}
\newblock
\APACjournalVolNumPages{IEEE Software}{33}{6}{70--79}.
\PrintBackRefs{\CurrentBib}

\bibitem [\protect \citeauthoryear {%
Thoring%
, Desmet%
\BCBL {}\ \BBA {} Badke-Schaub%
}{%
Thoring%
\ \protect \BOthers {.}}{%
{\protect \APACyear {2019}}%
}]{%
thoring2019creative}
\APACinsertmetastar {%
thoring2019creative}%
\begin{APACrefauthors}%
Thoring, K.%
, Desmet, P.%
\BCBL {}\ \BBA {} Badke-Schaub, P.%
\end{APACrefauthors}%
\unskip\
\newblock
\APACrefYearMonthDay{2019}{}{}.
\newblock
{\BBOQ}\APACrefatitle {Creative space: a systematic review of the literature}
  {Creative space: a systematic review of the literature}.{\BBCQ}
\newblock
\BIn{} \APACrefbtitle {Proceedings of the Design Society: International
  Conference on Engineering Design} {Proceedings of the design society:
  International conference on engineering design}\ (\BVOL~1, \BPGS\ 299--308).
\PrintBackRefs{\CurrentBib}

\end{thebibliography}


\end{document}